\documentclass{PoS}

\title{Highlights from galactic observations with MAGIC}

\ShortTitle{MAGIC galactic observations}

\author{\speaker{Marcos L\'opez} for the MAGIC Collaboration \thanks{https://magic.mpp.mpg.de}\\
        Univ. Complutense de Madrid, Fac. C.C. F\'isicas and UPARCOS, 2840 Madrid, Spain\\
        E-mail: \email{marcos@gae.ucm.es}}

\abstract{MAGIC is one of the main instruments for exploring the galactic gamma-ray sky from ground in the energy range of 50 GeV - 50 TeV. It consists of two 17 m diameter imaging atmospheric Cherenkov telescopes located at the Roque de los Muchachos Observatory, on the Canary island of La Palma. Thanks to its 
excellent sensitivity, MAGIC has conducted relevant studies on galactic objects of different types at Very High Energies. 
Among them, the  Crab pulsar up to TeV energies,  the spectral cut-off of the supernova remnant Cassiopeia A, the super-orbital variability of the binary system LSI +61 303, the search for microqusars, multi-year observations of the Galactic Center and follow-up studies of  unidentified HAWC sources.
In many cases, the results from these observations challenge our understanding of the underlying emission mechanisms.
Here we  review the latest results from the observation of these galactic sources  with MAGIC. 
}

\FullConference{35th International Cosmic Ray Conference --- ICRC2017\\
		10--20 July, 2017\\
		Bexco, Busan, Korea}

\begin{document}

%%%%%%%%%%%%%%%%%%%%%%%%%%%%%%%
%
%
\section{The MAGIC Telescopes}
The two 17 meter diameter MAGIC telescopes constitute a 
current generation instrument for Very High Energy (VHE) $\gamma$-ray observations,
exploiting the Imaging Air Cherenkov (IAC) technique. 
The telescopes are located  at an altitude
of 2200 m a.s.l. on the Roque de los Muchachos Observatory, in
La Palma island (Spain).
 MAGIC detects the faint flashes of Cherenkov light produced when  
 $\gamma$-rays (or cosmic-rays) plunge into the earth atmosphere and initiate
 showers of secondary particles. The Cherenkov light emitted by the charged
 secondary particles is reflected by the mirrors of the telescopes and an image
 of the 
 shower is obtained in each telescope camera. An offline analysis of the shower
 images allows the rejection of the hadronic 
cosmic ray background, the measurement of the direction of the incoming 
$\gamma$-rays, and the estimation of their energy. 

The first MAGIC telescope started observations in 2004, incorporating a number
of technological improvements in its design. 
The introduction of a second telescope in fall 2009, enabled
the instrument to perform stereoscopic observations 
with significantly better sensitivity and angular resolution.  
The energy threshold reached in stereoscopic mode is as low as 50 GeV, and the system achieves an integral sensitivity of $0.66\%$ of the Crab Nebula flux in 50 hours of observation above 220 GeV. 
For more details about the current
instrument's performance, please see  \cite{CrabPerformance}. 

%%%%%%%%%%%%%%%%%%%%%%%%%%%%%%%
%
%
\section{Pulsars and Pulsar Wind Nebulae}
\begin{figure}
\centering
\includegraphics[width=.52\textwidth]{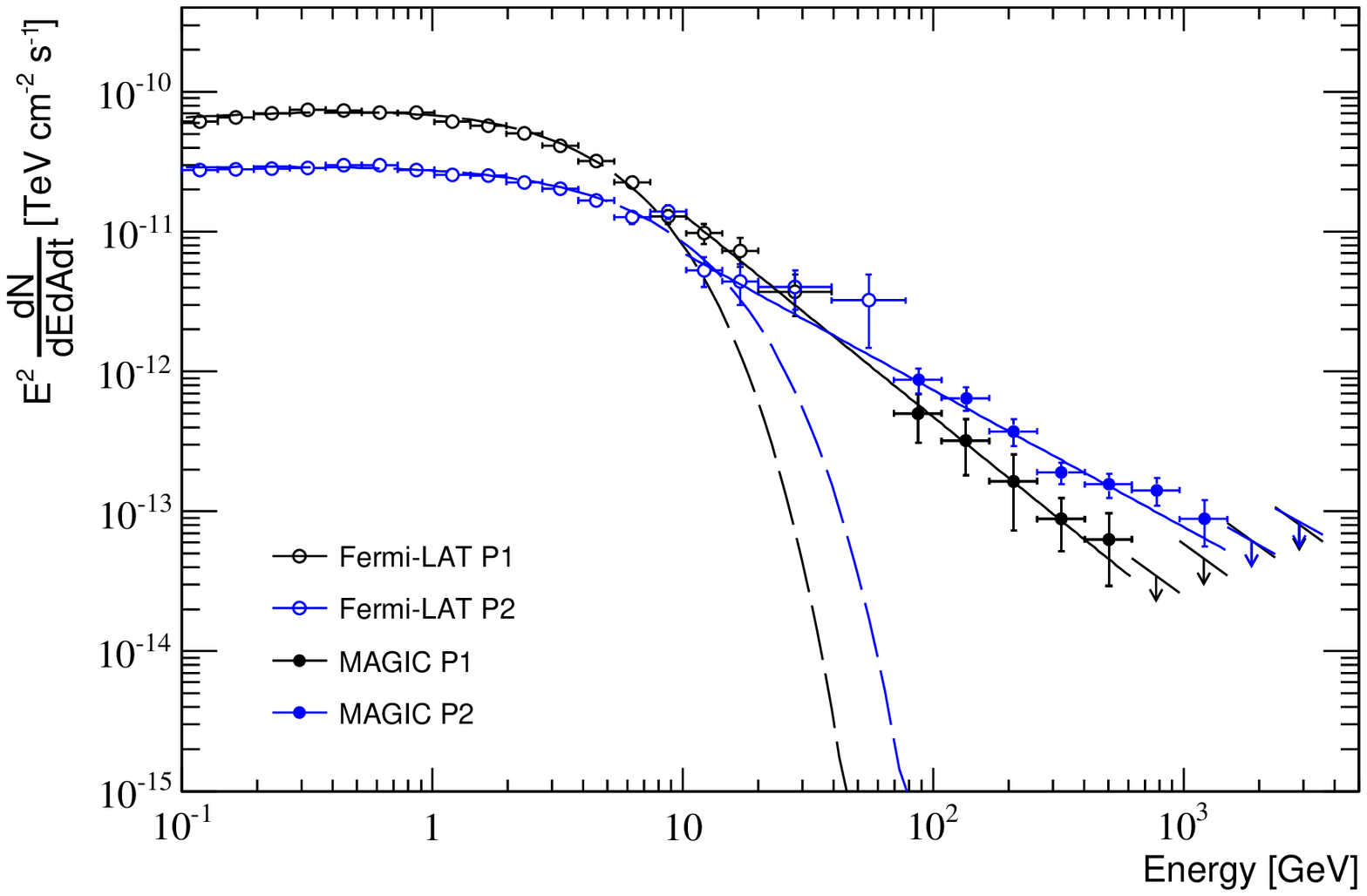}
\includegraphics[width=.47\textwidth]{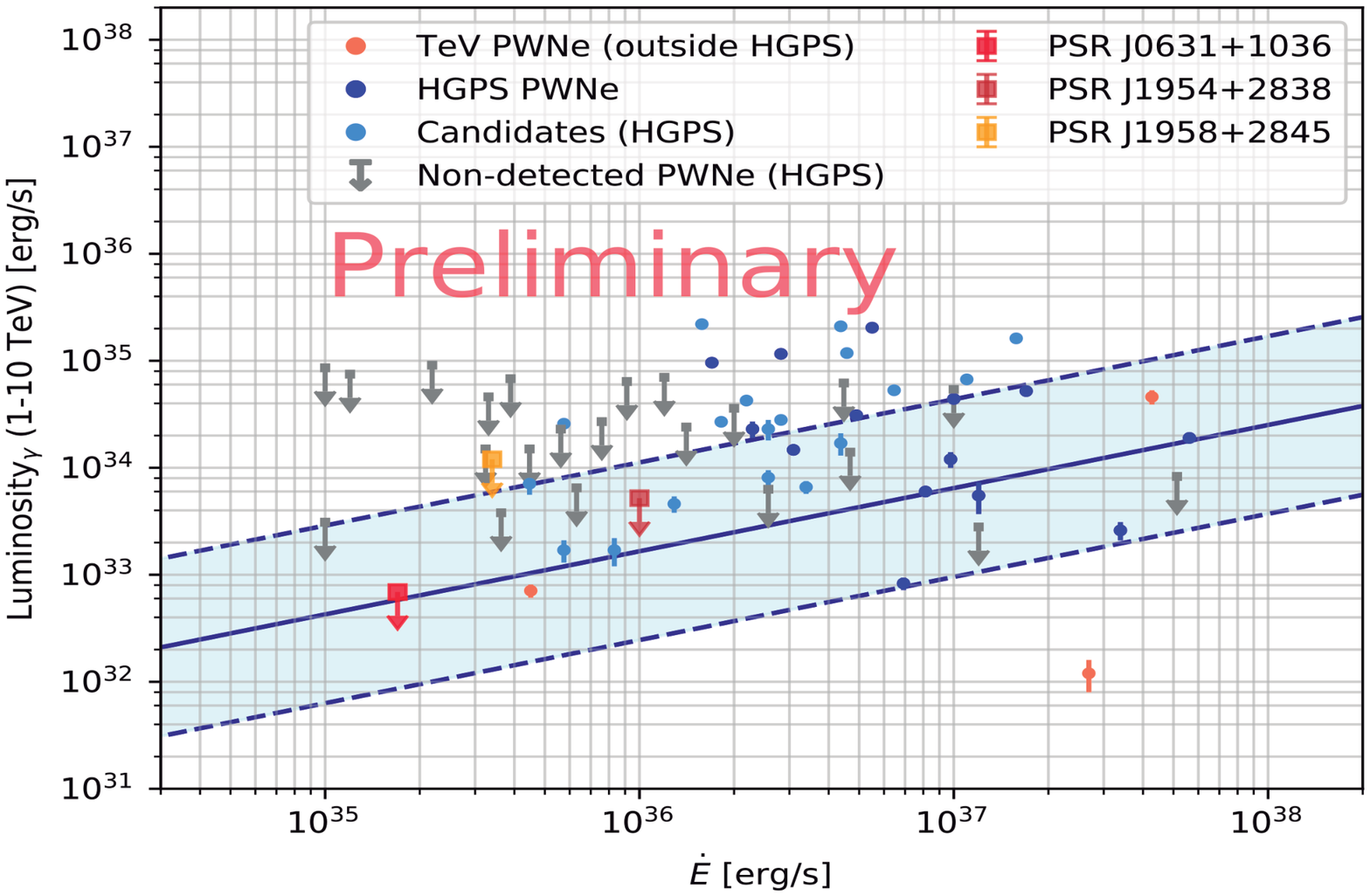}
\caption{{\bf Left}: 
Phase-folded SED of the Crab pulsar emission peaks P1 (filled black circles) and P2 (filled blue circles) measured by MAGIC between $\sim$70\,GeV and 1.5\,TeV. The results of the power-law with exponential cut-off fits to the Fermi-LAT points (open circles) are illustrated by the dashed lines, whereas the joint Fermi-LAT/MAGIC fits to power-law functions above 10 GeV are shown by solid lines. 
{\bf Right}: 
TeV luminosity (1 - 10 TeV) with respect to the spin-down power of the pulsar obtained in the study of PWNe by \cite{HGPS}. The best fit 
is depicted as a blue band. The 3 candidates observed by MAGIC are marked with squares. }
\label{fig1}
\end{figure}

The mechanism of the pulsed electromagnetic emission in the Crab pulsar is still an open
fundamental question. Models for High Energy (HE) emission 
predict exponential
or super-exponential cut-offs in pulsar spectra at a few GeV, in agreement to what Fermi-LAT
is measuring for all pulsars in the 100 MeV - 10 GeV energy range. 
However, MAGIC has detected pulsed emission from
the Crab pulsar above 25 GeV \cite{Crab2008}. 
The latest MAGIC publication combines more than 7 years of Crab pulsar data, from 2007 to 2014, resulting in 320 hours of observation time \cite{CrabTeV}. It is worth noting that this data set constitutes one of the deepest observation of any particular object in the VHE regime performed by IACTs to date.
With this exceptional data set MAGIC has been able to detect for the first time pulsed emission above 400 GeV and revealed the extension of Crab's VHE tail to 1.5 TeV without any hint of a cut-off (see Fig. \ref{fig1} left).
The combination of MAGIC's phase resolved measurements with results obtained by the Fermi-LAT, showed that the VHE tail is significantly harder for P2 than for P1. Hence, P2 solely dominates the pulse profile above 400 GeV, with P1 being visible only at a 2.2 $\sigma$ level.
Emission of pulsed TeV photons strongly favors inverse Compton scattering over synchro-curvature radiation as the responsible emission mechanism. It is still under debate where in the magnetosphere of the pulsar the up-scattering takes place and if the emission zone lies within or outside the light cylinder. Other open issues that have yet to be addressed are the narrow peaks observed in the pulse profile above 100 GeV and their phase coherence along
the entire electromagnetical spectrum, from radio up to TeV energies. 

%\subsection{PWN}
Pulsar wind nebulae (PWNe) constitute the most numerous TeV gamma-ray sources
in our galaxy. Most of the 19 firmly identified pulsar wind nebulae to date were spotted in the H.E.S.S Galactic Plane Survey (HGPS) of the inner Milky Way from the southern hemisphere \cite{HGPSlink}. Also MAGIC contributed valuable case studies to this population with the discovery of the least luminous PWN 3C 58 \cite{3c58}, and by extensively studying the most luminous one, the Crab nebula \cite{CrabNebula2015}. TeV pulsar wind nebulae are usually located close to a galactic spiral arm structure, where the dense environment provides a natural birth place for pulsars. In particular, the Scutum-Centaurus arm near the Galactic center hosts almost half of the current TeV PWNe population \cite{HGPS}. 
To increase the population with members situated towards the outer part of our galaxy, MAGIC selected and observed 3 young and energetic gamma-ray pulsars that could potentially power TeV pulsar wind nebulae: PSR J0631+1036, PSR J1954+2838 and PSR J1958+2845 \cite{AlbaPWNICRC2017}. For all 3 candidates the reanalysis of the full eight-year Milagro data set showed hot spots at the $\sim 4 \sigma$ level \cite{MilagroPWN}. However, the observations conducted by MAGIC did not reveal significant excess in the direction of any of the candidates above 300 GeV. In the case of PSR J0631 the obtained integral upper limit (UL) corroborates HAWC's non-detection and favors the interpretation of Milagro's hot spot in terms of a statistical artifact. The integral ULs for all 3 candidates, once converted into luminosity UL and computed in the 1 to 10 TeV range, are in agreement with the observational correlations seen in the TeV PWNe population so far and should not significantly alter the fits obtained by H.E.S.S. \cite{HGPS} (see Fig.  \ref{fig1} right).

%%%%%%%%%%%%%%%%%%%%%%%%%%%%%%%
%
%
\section{Supernova Remnants: A cut-off in Cassiopeia A}
It is widely believed that Galactic Cosmic Rays (CR) are accelerated in Supernova Remnants
(SNRs) through the process of diffusive shock acceleration. In this scenario, particles should be
accelerated up to energies around 1 PeV (the so-called 'Knee') and emit $\gamma$ rays. To test
this hypothesis precise measurements of the $\gamma$-ray spectra of young SNRs at TeV energies
are needed. Among the already known SNRs, Cassiopea A (Cas A) appears as one of the
best candidates for such studies, because it is relatively young (about 300 years) and it has been
largely studied in radio and X-ray bands, which constrains essential parameters for testing emission
models. 
With this goal in mind MAGIC has conducted a multi-year observation campaign of Cas A, between December 2014 and October 2016, for a total of 158 hours \cite{MAGIC_CasA}. 
This deep observation has allowed the most precise spectrum of Cas A to date, from 100 GeV to 9 TeV. The spectrum is best described assuming a power-law distribution with an exponential cut-off at 3.5 TeV. The cut-off fit is preferred with 4.6 $\sigma$ over a pure power-law scenario. This is the first observational evidence in the VHE regime of a spectral cut-off in Cas A.
Assuming that TeV $\gamma$-rays are produced by
hadronic processes and that there is no significant cosmic ray diffusion, this indicates that Cas A
is not a PeVatron (PeV accelerator) at its present age.

%%%%%%%%%%%%%%%%%%%%%%%%%%%%%%%
%
%
\section{Binary systems}
\begin{figure}
\centering
\includegraphics[width=.6\textwidth]{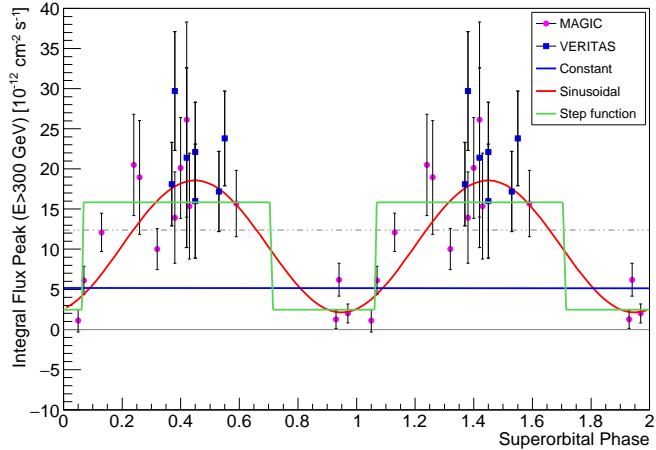}
\caption{Peak of the VHE emission in terms of the super-orbital phase defined in radio  \cite{LSI2016}. Each data point represents the peak flux emitted in one orbital period during orbital phases 0.5 - 0.75 and is folded into the super-orbit of 1667 days known from radio observations. MAGIC (magenta dots) and VERITAS (blue squares) points have been used in this analysis. The fit with a sinusoidal (solid red line), with a step function (solid green line), and with a constant (solid blue line) are also represented. The gray dashed line represents 10\% of the Crab Nebula flux, the gray solid line the zero level for reference.}
\label{fig2}
\end{figure}

LS I $+$61 303 is a $\gamma$-ray binary composed of a rapidly rotating Be star  with a circumstellar disk and a compact object of unknown nature, either a neutron star (NS) or a stellar-mass black hole (BH). %, with an orbital period of 26 days. 
After the discovery of LS I+61 303 in 2006 \cite{LSI2006}, MAGIC has continued observing the source. 
The VHE emission decreased significantly in 2008, but, thanks to the increased sensitivity of the stereo system, MAGIC detected it  at a flux of less than 5\% of the Crab flux during this state of low emission in 2009.
As part of a multi-wavelength campaign, MAGIC observed again the source between August 2010 and September 2014 \cite{LSI2016}. 
All archival data of LS I +61 303 recorded by MAGIC since its detection in 2006  were folded onto the superorbital period of 1667 days (see Fig. \ref{fig2}). The data were fit with a constant, a sinusoidal and with two-emission levels (step function).
The probability for a constant flux is negligible. Assuming a sinusoidal signal, the fit probability reaches 8\%, while the fit to a step function resulted in a fit probability of 7\%. This shows that the observed intensity distribution can be described
by a high and a low state and with a smoother transition. We conclude that there is a super-orbital signature in the TeV emission that it is compatible with the 4.5-year radio modulation seen in other frequencies.

\subsection{Galactic jets}
Conversely to the $\gamma$-ray binaries, there is no firm evidence  so far of VHE emission from microquasar jets. 
Gamma-ray observations of microquasars at VHE energies can provide valuable
information of the acceleration processes inside the jets, the jet-environment interaction and the
disk-jet coupling. MAGIC  has been deeply studying two high-mass microquasars to shed light on these
aspects: Cygnus X-1 and Cygnus X-3. Both systems display the canonical hard and soft X-ray
spectral states of black hole transients, where the radiation is dominated by non-thermal emission
from the corona and jets and by thermal emission from the disk, respectively.
 
MAGIC monitored Cygnus X-1 for 50 h in 2006, and
found an evidence of signal at the level of 4.1$\sigma$ 
on September 24th, 2006 \cite{CygX1}. The source was in a high state of emission
and in coincidence with a hard X-ray flare. Following this
promising result, MAGIC observed the source for an additional
100 hours between July 2007 and November 2009, at different
X-ray states \cite{AlbaCygnusICRC2017}. No significant excess was found above 200 GeV. Orbital phase-folded and daily analysis were also carried out, with no evidence of emission.
Although VHE radiation is predicted in the models, e.g. \cite{Romero15}, several factors can prevent detection: low flux below MAGIC sensitivity, no efficient acceleration on the jets or strong magnetic fields. Nevertheless, transient
events related e.g. to discrete radio-emitting-blobs 
cannot be discarded, which could explain the hint of  emission reported  by MAGIC in 2006 \cite{CygX1}.

During summer 2016, Cygnus X-3 underwent a flaring activity
period in radio and HE $\gamma$-rays, similar to the one that led to its detection in the HE regime in 2009 \cite{CygnusX3Fermi2009}. MAGIC performed comprehensive follow-up observations for a total of 70 hours \cite{AlbaCygnusICRC2017}. 
No excess was found at energies above 100 GeV. No orbital  or daily modulation was detected either.
This  reinforces the idea that VHE $\gamma$ rays, if produced, are originated inside the
binary system and not at the radio-emitting regions of the jets far from the compact object.

%%%%%%%%%%%%%%%%%%%%%%%%%%%%%%%
%
%
\section{Galactic Center observations}
The central region of our galaxy is very densely populated with
a large variety of astrophysical objects, many of which may
be sites of extreme particle acceleration and hence $\gamma$-ray emission.
Among those, the compact radio
source Sagittarius A* (Sgr A*) is of a particular interest, and
is generally accepted to be associated with the $4\times 10^{6}$ \(\textup{M}_\odot\) super-massive black hole (SMBH) at the center of our galaxy. Several models for the production of
high-energy radiation from Sgr A* itself have been proposed, including
leptonic, hadronic and hybrid scenarios (see e.g. \cite{GCLeptonic, GCHadronic}).

\begin{figure}
\centering
\includegraphics[width=.6\textwidth]{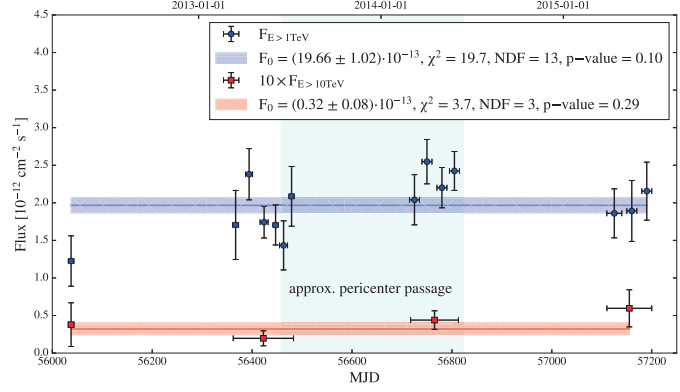}
\caption{Light curves obtained with MAGIC of the integral gamma-ray flux from the Galactic Center for E\,>\,1 TeV and E\,>\,10 TeV  \cite{MagicGCpaper}. The red and blue lines represent the
best fits to the constant flux, the corresponding shaded regions represent the 1$\sigma$ confidence intervals. The flux values for E\,>\,10 TeV have been multiplied
by 10 for better visibility in the plot.}
\label{fig3}
\end{figure}

Between 2012 and 2015, MAGIC has conducted a multi-year monitoring campaign of the Galactic Center (GC), collecting 67 hours of good-quality data  \cite{MagicGCpaper}. These
observations were primarily motivated by reports that a putative gas cloud (G2) would be passing in close proximity to Sag A*. This event was expected to give astronomers a unique chance to study the effect of in-falling matter on the broad-band emission of a SMBH.
No effect of the G2 object on the VHE $\gamma$-ray emission
from the GC was detected during the 4 year observation campaign (see Fig. \ref{fig3}).
The lack of variability from the direction of Sgr A*, as
measured by MAGIC, makes it difficult to rule out single models
describing particle acceleration and $\gamma$-ray emission mechanisms
at the source. 
Nevertheless, these observations may still prove useful
as an accurate measurement of the baseline emission from Sgr A* in the case of any possible flaring activity in the future.
Along with the variability study, the large exposure allowed MAGIC to derive a precise energy spectrum of Sgr A*,
which agrees with previous measurements within errors. A study of the morphology of the GC region has led to the detection of
 a VHE source of unknown nature in the region of the GC Radio Arc, 
 and to the confirmation of the supernova remnant G0.9+0.1 as a VHE emitter  \cite{MagicGCpaper}.
 
%%%%%%%%%%%%%%%%%%%%%%%%%%%%%%%
%
%
\section{Follow-up studies of HAWC sources}
The HAWC observatory, inaugurated in 2015, is sensitive to cosmic and $\gamma$ rays in the energy range from 100 GeV to 100 TeV.
The second HAWC catalog (2HWC) \cite{HAWC}, released in 2017,  contains 39 detected TeV sources, from which 19 of them have no association with any known VHE source.
The 2HWC catalog motivated follow-up studies with MAGIC. The aim was to focus on the  sources
with no association, in order to provide new information of unknown candidates. After evaluating those sources, a short list of 3 candidates were selected: 2HWC J2006$+$341, 2HWC J1907$+$084 and
 2HWC J1852+013. All these candidates were located in the FoV of former MAGIC observations, 
within $ < 1.5^{\circ}$ from the nominal observation position, allowing MAGIC to analyze these sources without performing new dedicated observations. The 3 candidates were detected during the HAWC
point-like search and therefore, they might be point-like sources for MAGIC as well. 
The total data samples of MAGIC covering these sources amounts to 61, 4 and 120 hours respectively, recorded in different period between 2013 and 2016.
 We did not detect any of the candidates with MAGIC or any hotspot nearby, neither assuming a point-like source  ($0.10^{\circ}$) or slightly extended one ($0.16^{\circ}$). 
However, MAGIC observations allow us to set constrains on the extension of the sources. For 2HWC J2006+341, an assumption of
$\sim 0.16^{\circ}$ in MAGIC data seems to be already compatible with HAWC results. 
In the cases of 2HWC J1907+084 and 2HWC J1852+013, 
MAGIC and HAWC results are in agreement in the TeV regime. 

Given that the largest population of TeV emitters in our Galaxy are the Pulsar Wind Nebulae (PWNe), it would not be improbable that the 3 selected candidates belong to this type of sources. However, a search for detected pulsars nearby the 2HWC sources using the ATNF catalog \cite{ATNFpulsars} reveals that none of the detected pulsars can be related with these objects, questioning a possible PWN nature.

%%%%%%%%%%%%%%%%%%%%%%%%%%%%%%%
%
%
\section{Stereo SumTrigger system}

Gamma-ray observations from ground are  limited by the overwhelming background at lower energies. 
The analog SumTrigger concept is an alternative to the  standard digital trigger, providing a better background suppression, and hence a lower trigger threshold. 
In 2007 a prototype of an analog SumTrigger was installed in the MAGIC-I telescope, which lowered the trigger threshold from 55 GeV to 25 GeV and led to the first detection of the Crab pulsar at VHE \cite{Crab2008}. 
In the latest years a new version of the system has been developed to be used during stereoscopic observations. 
 An advantage of the new system is that the stereoscopic implementation partially does the work of the clipping stage \cite{SumTrigger}, since the probability that huge after-pulses occur at the same time in both telescopes is reduced by two orders of magnitude
The new system has been installed on both MAGIC telescopes in 2014. 
The analysis of the first Crab data taken with the new  system reveals that at low energies the Stereo SumTrigger 
outperforms the standard digital trigger, providing $40\%$ more pulsed photons.

\section{ Summary}
For the last 13 years the MAGIC telescopes have contributed to improve our knowledge of the Galactic sky at TeV energies. The latest highlights are the detection of the Crab pulsar up to TeV energies, the measurement of a spectral cut-off in Cas A and the discovery of a  super-orbital signature in the TeV emission of LSI +61 303. In the meantime, MAGIC has continued developing new technologies for pushing down the energy threshold, as with the development of the stereo SumTrigger system, which will allow further discoveries in the coming years.

\section*{Acknowledgements}
We would like to thank the IAC for the excellent working conditions at the ORM in La Palma. We acknowledge the financial support of the German BMBF, DFG and MPG, the Italian INFN and INAF, the Swiss National Fund SNF, the European ERDF, the Spanish MINECO, the Japanese JSPS and MEXT, the Croatian CSF, and the Polish MNiSzW.

\end{document}